\documentclass{article}
\usepackage{spconf,amssymb,amsmath,graphicx}
\usepackage{epsfig}
\usepackage{xspace}
\usepackage{srcltx}
\usepackage{caption}
\usepackage{subcaption}
\usepackage[export]{adjustbox}
\usepackage{multirow,makecell,hhline}
\usepackage{booktabs}
\usepackage{textcomp}
\usepackage{cite}

\usepackage{url}
\usepackage{hyperref}
\hypersetup{colorlinks=True, urlcolor=black}

\usepackage{color}
\usepackage[dvipsnames]{xcolor}

\newcommand{\blue}[1]{{\color{blue}#1}}

\newcommand{\CMT}[1]{{}}

\newcommand{\tbh}[1]{\textbf{#1}}

\title{A Better and Faster End-to-End Model for Streaming ASR}
\name{Bo Li\textsuperscript{*}, Anmol Gulati\textsuperscript{*}, Jiahui Yu\sthanks{Equal contribution}, Tara N. Sainath, Chung-Cheng Chiu, Arun Narayanan,}
\nameplus{Shuo-Yiin Chang, Ruoming Pang, Yanzhang He, James Qin, Wei Han, Qiao Liang, Yu Zhang,}
\nameplusplus{Trevor Strohman, Yonghui Wu}
\address{Google LLC, USA \\
\fontsize{9}{9}\selectfont\ttfamily\upshape
\{boboli,anmolgulati,jiahuiyu\}@google.com}
\begin{document}
\ninept
\maketitle
\begin{abstract}
End-to-end (E2E) models have shown to outperform state-of-the-art conventional models for streaming speech recognition\cite{sainath20streaming} across many dimensions, including quality (as measured by word error rate (WER)) and endpointer latency\cite{li2020towards}. However, the model still tends to delay the predictions towards the end and thus has much higher partial latency compared to a conventional ASR model. To address this issue, we look at encouraging the E2E model to emit words early, through an algorithm called \emph{FastEmit}\cite{yu21fastemit}. Naturally, improving on latency results in a quality degradation. To address this, we explore replacing the LSTM layers in the encoder of our E2E model with \emph{Conformer} layers\cite{gulati2020conformer}, which has shown good improvements for ASR. Secondly, we also explore running a 2nd-pass beam search to improve quality. In order to ensure the 2nd-pass completes quickly, we explore non-causal Conformer layers that feed into the same 1st-pass RNN-T decoder, an algorithm called \emph{Cascaded Encoders}\cite{arun21cascade}. Overall, the Conformer RNN-T with Cascaded Encoders offers a better quality and latency tradeoff for streaming ASR.
\end{abstract}

\vspace{-0.05in}
\section{Introduction} 
\label{sec:intro}
\vspace{-0.05in}

Having an end-to-end (E2E) speech recognition model \cite{li2020comparison,Ryan19,CC18,Graves12,Chan15,KimHoriWatanabe17}, that matches/surpasses the quality of a conventional model (which has separate acoustic, pronunciation and language model components) across all dimensions of quality and latency, has become a challenging research direction over the past few years. 

On the latency side, one such requirement of E2E models is that they have to be streaming. They must emit words to the screen with minimal delay as the user speaks. Models such as Recurrent Neural Network Transducer (RNN-T) \cite{Graves12}, Neural Transducer \cite{Jaitly16, tsunoo2020streaming}, Monotonic Attention \cite{colin17,fan2018online}, Triggered attention \cite{moritz2019triggered} and Scout Network \cite{wang2020low} have been explored to address the streaming nature. Another requirement is that the model must stop decoding quickly once the user stops speaking, a problem known as endpointing. Folding the endpointer into the E2E model helped address this issue \cite{li2020towards}. 

On the quality side, we require a model which achieves better word error rate (WER) compared to a conventional model. Using a 1st-pass streaming RNN-T model coupled with a 2nd-pass non-streaming attention-based model helped address quality concerns \cite{SainathPang19}. In addition, improving the performance of the model to recognize proper nouns was achieved with contextual biasing \cite{Ding19}. 

Putting many of these above components together, we recently released an on-device model that surpassed a conventional model in terms of both WER and endpointer latency \cite{sainath20streaming}. However, one of the issues with that model is that partial hypotheses were still delayed to the end compared to a conventional model trained with an alignment. 

In this work, we explore ideas to improve the model's partial latency. Speech-to-text alignment can be used to improve the token prediction timing \cite{sainath20emitting,inaguma2020minimum}, which normally causes large quality degradation. We adopted the recently proposed FastEmit\cite{yu21fastemit}. It regularizes the RNN-T lattice to encourage non-blank tokens and surpass the blank token across the entire sequence based on the forward-backward probabilities. It is found to be effective in reducing models' partial latency with much less quality degradation. 

One can expect that any research attempt to improve partial latency results in a slight quality degradation. To address this, we explore improving our core E2E model by replacing the recurrent LSTM layers in the encoder with Conformer layers, which have shown promise in Librispeech \cite{gulati2020conformer}. In addition, we also look at improving our 2nd-pass strategy so that we can run a beam search rather than a rescorer \cite{arun21cascade}.  
We explored the improved latency and quality ideas on a Voice Search task, which yielded 10\% relative WER reduction with 250ms and 170ms reduction in median partial and prefetch latency respectively than \cite{sainath20streaming}.

\vspace{-0.05in}
\section{Model Improvements}
\label{sec:quality}
\vspace{-0.05in}

\subsection{Conformer}
\vspace{-0.05in}

We use Conformer~\cite{gulati2020conformer} to improve the model quality. The model used in \cite{gulati2020conformer} is designed for tasks with non-streaming, single-domain and short utterances. In this work, we adapt it for our streaming multi-domain task \cite{sainath20streaming}. We first limit the self-attention, the convolution and the normalization layers to take in only the previous context for streaming applications. We also replaced the full context self-attention with local self-attention which helps generalize to long utterances~\cite{chiu2019comparison}. The same as \cite{sainath20streaming}, we train our systems on a combination of data from different domains. For efficiency, we aggregate utterances with similar lengths before creating batches ({\it i.e.} bucketing). Due to the different length distributions across domains, for a particular batch the data may bias towards certain domain. This could lead to biased statistics for the batch normalization used in the original Conformer layer, and then limit the model's generalization capability across domains. We addressed this issue by replacing the batch normalization layer with the group normalization layer~\cite{wu2018group}. This is found critical for the quality of our multi-domain models. 

\begin{figure}[t]
\centering
    \includegraphics[scale=0.275]{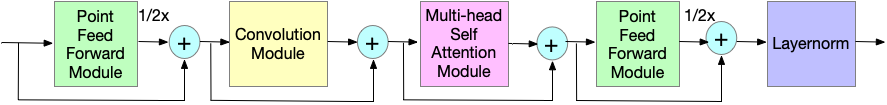}
    \caption{Improved Conformer block architecture.}
    \label{fig:conformer}
   \vspace{-0.2in}
\end{figure}

Additionally, we speedup the training and inference by removing the original relative positional encoding~\cite{dai2019transformerxl}. Instead, we reuse the existing convolution module that aggregates information from neighboring context to implicitly provide relative positional information~\cite{dai2019transformerxl,wang2019transformer}. This is done by simply swapping the order of the convolution module and the multi-head self-attention module, leading to the improved Conformer block depicted in Figure~\ref{fig:conformer}.

\subsection{Two-pass with Cascaded Encoders}
\vspace{-0.05in}

A big goal of two-pass models is to make up for the quality degradation in the 1st-pass. To improve our current two-pass approach, which is to run rescoring, we look to run a beam search instead. One of the challenges with this beam search is that we need the model to be robust to decoding short search utterances as well as long caption utterances. Typically attention-based models do very poorly on long-form \cite{chiu2019comparison}.

\begin{figure}[h!]
  \centering
  \includegraphics[scale=0.08]{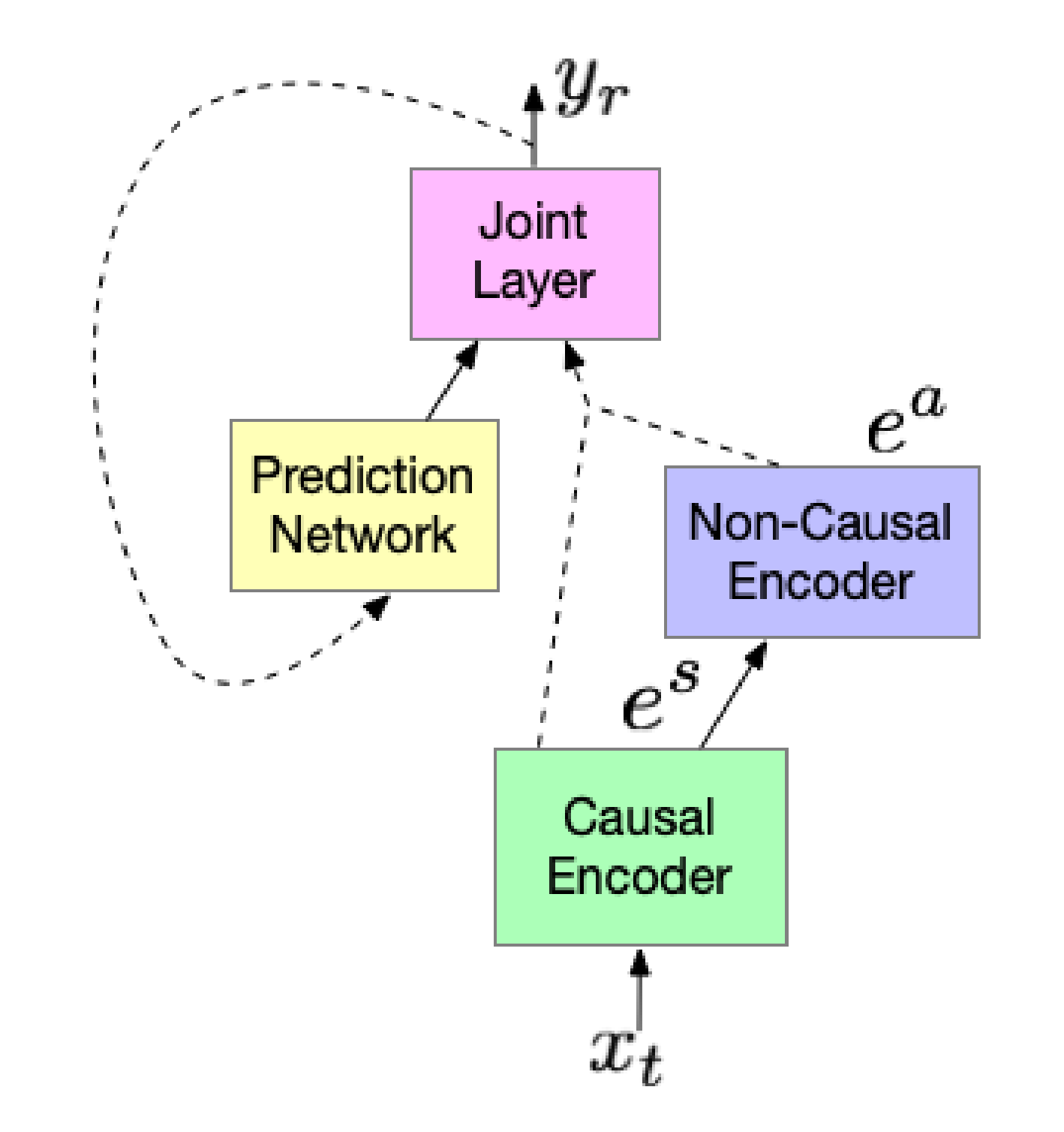}
  \caption{A block diagram of the Cascaded Encoders model.}
   \label{fig:cascade}
   \vspace{-0.2in}
\end{figure}

To address this, we look at running a 2nd-pass beam search with RNN-T itself, which has shown better robustness to long-form\cite{Arun19}. The model explored, which we call \emph{Cascaded Encoders}\cite{arun21cascade}, is shown in Figure \ref{fig:cascade}. To add the non-causal aspect of LAS (which improves quality)\cite{SainathPang19} into a streaming, causal RNN-T system, we add additional non-causal encoder layers on top of the causal encoder layers. The 1st-pass uses only the causal encoder and the RNN-T decoder. In the second pass, the additional non-causal layers take in both left and right context of the 1st-pass encoder outputs, and again feeds to the same decoder. A single RNN-T decoder is shared between the 1st and 2nd-passes for smaller model size and on-device benefits.

\section{Latency Improvements}
\label{sec:latency}
\vspace{-0.05in}

In this section, we present latency metrics for streaming ASR and techniques to reduce them. Different from the average token latency \cite{inaguma2020minimum}, we adopt metrics that are directly related to streaming speech applications such as Voice Search and Assistant.

\subsection{Endpointer Latency}
\vspace{-0.05in}

In streaming ASR based applications, when the user finishes speaking is a critical signal that triggers the next step fulfillment. For Voice Search/Assistant, it signals the ASR system to finalize the recognition result and initiate either a server request for more information or an action on the client side. \emph{Endpointer latency} measures the time difference between when the user finishes speaking and when the system predicts the end of the query (EOQ). This is the latency metric we normally referred to in previous work \cite{li2020towards,sainath20streaming}. A lower Endpointer latency leads to a lower application latency and faster system responses. We adopt the same RNN-T Endpointer model together with early and late penalties on EOQ during training to reduce Endpointer latency \cite{li2020towards}. We report both the median (i.e. $50^{th}$ percentile, EP50) and the $90^{th}$ percentile (EP90) latency for comparisons.

\subsection{Prefetch Latency}
\vspace{-0.05in}

Prefetching is a common technique used in web-based applications to reduce  network latency involved in the application stack \cite{luiz2003web,jonassen2012prefetching}. Instead of waiting for the EOQ signal, streaming applications normally send early server requests (either retrieving relevant information or necessary resources for next step actions) based on partial/incomplete recognition hypotheses. If a prefetch is based on a partial hypothesis that does not match the finalized hypothesis at the time EOQ is generated, it is discarded as a premature prefetching, increasing the overall computation cost. However, if the partial hypothesis used in a prefetching matches with the final result, the system can immediately execute the following action without waiting for additional server responses. This saves the time used to compute and fetch responses from the server and reduces the network processing latency. \emph{Prefetch latency} is defined as the time difference between when the first correct prefetch is triggered and when the user finishes speaking \cite{chang20low}. A prefetch is correct only if the partial recognition result it used matches the final recognition result. If no correct prefetch is found, the latency falls back to Endpointer latency. Besides latency, the prefetch rate, which is the number of prefetches for each query, is also an important metric. For comparison, we report both the $50^{th}$ and $90^{th}$ percentile latency (PF50 and PF90) together with the prefetching rate (PFR). 

\subsubsection{E2E Prefetching}
\vspace{-0.05in}

There are multiple ways to trigger prefetching \cite{chang20low}. It normally relies on observing a fixed interval of silence either from the decoding or an external voice activity detector (VAD). An E2E Prefetching has recently been proposed and yielded a much better prefetch latency and prefetch rate tradeoff \cite{chang20low}. It leverages on the EOQ prediction of the RNN-T Endpointer model for prefetching decisions. Different from endpointing, we want to generate EOQ for every partial result but avoid changing decoding. To achieve that, for each frame before EOQ shows on the top beam, we artificially add it to the end of the current hypothesis to compute its probability given the observations and the best path so far. This allows us to predict how likely the query is completed at each time frame and the corresponding partial result while decoding following frames. If the probability is greater than a predefined threshold, the system declares a prefetch decision.

\subsection{Partial Latency}
\vspace{-0.05in}

Prefetching relies on partial results generated by the ASR model. Systems that output partial hypothesis with less latency are preferred. We hence define \emph{partial latency} as the time difference between when the first correct partial hypothesis is generated by the model and when the user finishes speaking. Similarly, the correctness of a partial hypothesis refers to whether it matches the finalized recognition hypothesis. Unlike prefetching latency, partial latency is inherent to the ASR model itself and does not rely on additional logic implemented in the application pipeline such as the decision to trigger prefecthing. Partial latency is the lower bound for prefetch latency. Developing techniques that  reduce models' partial latency would give more room for reducing the overall application latency and is independent of the actual application pipeline.

\subsubsection{Constrained Alignment}
\vspace{-0.05in}

One way to improve partial latency is to constrain the model's prediction based on alignments generated from other models \cite{sainath20emitting}. Similar to the RNN-T EP model that constrains the EOU token, constrained alignment penalizes the prediction of each word's beginning and ending tokens if they fall out of a buffer centered around the corresponding ground truth obtained from the alignment. It has been shown to yield more accurate prediction timing information and can be used to improve partial latency.

\subsubsection{FastEmit}
\vspace{-0.05in}

In conventional RNN-T optimization, emitting a non-blank token and the blank token are treated equally, as long as the log-probability of the target sequence is maximized. However, in streaming ASR systems, the blank token (i.e. output nothing) should be discouraged as it leads to higher latency. A simple and effective sequence-level emission regularization technique, \emph{FastEmit}\cite{yu21fastemit}, is hence developed. It encourages predicting non-blank tokens and suppresses blank tokens across the entire sequence based on transducer forward-backward probabilities. It is implemented by adding a regularization hyper parameter $\lambda_{\text{FastEmit}}$ to the original Equation 20 in \cite{Graves06} (please refer to \cite{Graves06} for detailed notion definitions):
\begin{equation}
\frac{\partial \mathcal{L}}{\partial \text{Pr}(k|t, u)} = -\frac{\alpha(t, u)}{\text{Pr}(\mathbf{y}^*|\mathbf{x})} 
\begin{cases}
    \blue{(1 + \lambda_{\text{FastEmit}})}\beta(t, u+1) \text{~if~} k = y_{u+1} \\
    \beta(t+1, u) \text{~if~} k = \varnothing\\
    0 \text{~otherwise.}
\end{cases} \nonumber
\end{equation}
Intuitively, FastEmit applies a ``higher learning rate'' ($\lambda_{\text{FastEmit}} \geqslant 0$) to the prediction of non-blank token when back-propagating into the streaming ASR network, while the prediction of blank token remains the same. It has been shown to give significant lower latency while retaining recognition accuracy on different RNN-T models. More importantly, FastEmit does not require any prior alignment information \cite{sainath20emitting, inaguma2020minimum} and has no additional training or serving cost.
\section{Experimental Setup}
\label{sec:exps}
\vspace{-0.05in}

\textbf{Model Architecture:} All models are trained using a 128D log-Mel feature frontend \cite{Arun19}. The features are computed using 32ms windows with a 10ms hop. Features from 4 contiguous frames are stacked to form a 512D input representation, which is further sub-sampled by a factor of 3. A 16D one-hot domain-id vector is then appended to form a 528D input vector as the inputs to the model. Following \cite{gulati2020conformer}, we use 512D Conformer layers in the encoder. Causal convolution and left-context attention layers are used for the Conformer layer to strictly restrict the model to use no future inputs. 8-head attention is used in the self-attention layer and the Convolution kernel size used is 15. The encoder consists of 17 Conformer layers. The RNN-T decoder consists of a prediction network with 2 LSTM layers with 2048 units projected down to 640 output units, and a joint network with a single feed-forward layer with 640 units. Two non-causal Conformer layers that take in additional 5.04s right context are used in the Cascaded Encoders on top of the streaming encoder before the decoder. All the E2E models are trained to predict 4,096 word pieces~\cite{Schuster2012}. The decoding is rescored with a 2nd-pass MaxEnt language model\cite{Biadsy17}.

The Conformer RNN-T model has 137M parameters. The two non-causal Conformer layers in the Cascaded Encoders model bring in additional 12M parameters. All models are trained in Tensorflow~\cite{AbadiAgarwalBarhamEtAl15} using the Lingvo~\cite{shen2019lingvo} toolkit on $8\times8$ Tensor Processing Units (TPU) slices with a global batch size of 4,096 utterances. Models are optimized using synchronized stochastic gradient descent with the Adam optimizer \cite{KingmaBa15}.

\noindent
\textbf{Datasets:} All the E2E models are trained on audio-text pairs the same as \cite{sainath20streaming}, which is a small fraction of data compared to the trillion-word text-only data a conventional LM is trained with. To increase vocabulary and diversity of training data, we explore using more data by incorporating multi-domain utterances as described in \cite{Arun19}. These multi-domain utterances span domains of search, farfield, telephony and YouTube. All datasets are anonymized and hand-transcribed; the transcription for YouTube utterances is done in a semi-supervised fashion ~\cite{liao2013large}. In addition to the diverse training sets, multi-condition training (MTR) ~\cite{kim2017mtr} and random data down-sampling to 8kHz \cite{Li12} are also used to further increase data diversity. Noisy data is generated at signal-noise-ratio (SNR) from 0 to 30~dB, with an average SNR of 12~dB, and with T60 times ranging from 0 to 900ms, averaging 500ms. Noise segments are sampled from YouTube and daily life noisy environmental recordings. Both 8~kHz and 16~kHz versions of the data are generated, each with equal probability, to make the model robust to varying sample rates. 

The test set includes around 12K Voice Search utterances with duration less than 5.5s long. They are anonymized and hand-transcribed, and are representative of Google's Voice Search traffic.

\vspace{-0.1in}
\section{Results}
\label{sec:results}
\vspace{-0.05in}

\begin{table}[t]
\caption{Performance of RNN-T models with different encoders.}
\centering
\begin{tabular}{l|c|c|c}
\toprule
\multirow{2}{*}{\tbh{Exp.}} & \tbh{Model} & \tbh{Training Speed} & \tbh{WER} \\
~ & \tbh{Size} \scriptsize{(M)} & \scriptsize{(examples/sec.)} & \scriptsize{(\%)} \\
\midrule
\midrule
\tbh{B1} \scriptsize{LSTM w MWER \cite{sainath20streaming}} & 122 & 3100 & 6.0 \\
\midrule
\tbh{C0} \scriptsize{w/o MWER \cite{gulati2020conformer}} & 141 & 3970 & 6.8 \\
\tbh{C0} \scriptsize{No bucketing, w/o MWER} & 141 & 2450 & 5.8 \\
\midrule
\tbh{C1} \scriptsize{w/o MWER} & 141 & 3550 & 5.9 \\
\midrule
\tbh{C2} \scriptsize{w/o MWER} & \multirow{2}{*}{137} & \multirow{2}{*}{4200} & 5.8 \\
\quad~ \scriptsize{w MWER} & ~ & ~ & 5.6 \\
\bottomrule
\end{tabular}
\label{tbl:conformer}
\vspace{-0.1in}
\end{table}

\begin{table*}[t]
\caption{Quality and Latency comparisons of different RNN-T models on the Voice Search test set.}
\centering
\begin{tabular}{l|c|cc|cc|ccc}
\toprule
\multirow{2}{*}{\tbh{Exp.}} & \tbh{WER} & \multicolumn{2}{c|}{\tbh{Endpointer Latency}} & \multicolumn{2}{c|}{\tbh{Partial Latency}}  & \multicolumn{3}{c}{\tbh{Prefetch Latency}}  \\
~ & \scriptsize{(\%)} & \scriptsize{EP50 (ms)} & \scriptsize{EP90 (ms)} & \scriptsize{PR50 (ms)} & \scriptsize{PR90 (ms)} & \scriptsize{PF50 (ms)} & \scriptsize{PF90 (ms)} & \scriptsize{PFR} \\
\midrule
\midrule
\tbh{B0} \scriptsize{Conventional \cite{sainath20streaming}} & 6.6 & 460 & 870 & -150 & 60 & 90 & 190 & 1.48 \\
\tbh{B1} \scriptsize{LSTM RNN-T\cite{sainath20streaming}} & 6.0 & 310 & 710 & 170 & 310 & 170 & 320 & 1.80 \\
\midrule
\tbh{B2} \scriptsize{B1 + Constrained Alignment} & 6.9 & 230 & 560 & -40 & 80 & 100 & 200 & 1.29 \\
\tbh{B3} \scriptsize{B1 + FastEmit} & 6.2 & 330 & 650 & -10 & 180 & 80 & 210 & 1.47 \\
\midrule
\tbh{C2} \scriptsize{Conformer RNN-T} & 5.6 & 260 & 590 & 150 & 290 & 220 & 350 & 1.65 \\
\tbh{C3} \scriptsize{C2 + FastEmit} & 5.8 & 290 & 660 & -110 & 90 & 70 & 210 & 1.29 \\
\tbh{C4} \scriptsize{C3 + E2E Prefetch} & 6.0 & 290 & 660 & -110 & 90 & -50 & 110 & 1.86 \\
\midrule
\tbh{T1} \scriptsize{Two-pass LAS Rescoring} & 5.3 & 290 & 660 & -100 & 140 & 80 & 210 & 1.30 \\
\midrule
\tbh{T2} \scriptsize{Single-pass Causal} & 6.0 & 290 & 660 & -90 & 120 & -20 & 130 & 1.90  \\
\tbh{T2} \scriptsize{Two-pass} & 5.4 & 290 & 660 & -80 & 140 & 0 & 140 & 1.84 \\
\tbh{T2} \scriptsize{Single-pass Non-causal} & 4.8 & - & - & - & - & - & - & -  \\
\bottomrule
\end{tabular}
\label{tbl:latency}
\vspace{-0.2in}
\end{table*}

\subsection{Quality Improvement with Conformer}
\vspace{-0.05in}

First we experimented with Conformer encoders for quality improvement. Table~\ref{tbl:conformer} presents the comparison on the Voice Search testset for RNN-T models with different encoders. The RNN-T model with an LSTM encoder serves as our baseline (B1), which has been shown to give a better quality and latency performance than conventional hybrid ASR systems \cite{sainath20streaming}. Limiting the Conformer encoder presented in \cite{gulati2020conformer} to use only left-context for streaming applications, system C0 shows significant WER degradation due to the biased batch normalization statistics in our multi-domain task. This can be resolved by randomly sampling utterances ({\it i.e.} no bucketing) instead of aggregating utterances with similar lengths into each batch (C0 No bucketing in Table~\ref{tbl:conformer}) with the cost of large training speed reduction due to the lower computation utilization per batch. By replacing the batch normalization with the group normalization, system C1 addressed the issue with similar WER gains but less training speed regression. Further swapping the order of the convolution module and the self-attention module, system C2 achieves both WER gains and training speedups. Comparing to B1, C2 yields a total of 7\% relative WER reduction and 35\% speedup. Note for B1 we did not see quality gains with more LSTM encoder layers.

\vspace{-0.05in}
\subsection{Latency Improvement}
\vspace{-0.05in}

In previous work \cite{sainath20streaming, li2020towards}, we focused on improving RNN-T models' endpointer latency and have shown that E2E models have a better WER and endpointer latency tradeoff than conventional hybrid ASR systems. However, even with that we still observe large total latency in E2E based streaming applications. This is then tracked down to the fact that the E2E systems tend to delay the partial hypotheses towards the end of the utterance where the endpointer fires, leaving less space for prefetching to reduce network latency (B0 vs. B1 in Table~\ref{tbl:latency}). 

\vspace{-0.05in}
\subsubsection{Reducing Partial Latency with Constrained Alignment}
\vspace{-0.05in}

In this experiment, we explored constraining the RNN-T prediction with alignment information following \cite{sainath20emitting}. When applied to B1, the use of constrained alignment achieves 210ms and 230ms latency reduction for PR50 and PR90 respectively with the cost of a quality degradation from 6.0\% to 6.9\% (B2 vs. B1 in Table~\ref{tbl:latency}). Negative latency suggests the model is capable of hypothesizing the whole query even before seeing the end of the speech.

\vspace{-0.05in}
\subsubsection{Reducing Partial Latency with FastEmit}
\vspace{-0.05in}

We then applied FastEmit to the same baseline system B1. From Table~\ref{tbl:latency}, B1 with FastEmit (B3) gives 180ms and 130ms reduction for PR50 and PR90 with much less WER regression (6.0\% to 6.2\%). Additionally, no alignment is needed during B3 training. To address the quality degradation, we used FastEmit on a better E2E model - C2 from Table~\ref{tbl:conformer}. Although FastEmit still brings in 0.2\% absolute WER regression, the system `C2+FastEmit' ({\it i.e.} C3) has a WER of 5.8\% with a -110ms PR50 and 90ms PR90, which is better both in quality and latency than B1. 

\vspace{-0.05in}
\subsubsection{Reducing Prefetch Latency with E2E Prefetching}
\vspace{-0.05in}

Next, we explored the use of E2E prefetch to minimize the gap between partial and prefetch as prefetch latency is the one that is reflected in the total latency. E2E prefetch can only be used in E2E systems and is not applicable to conventional models, where silence-based prefetching is used. From Table~\ref{tbl:latency}, adding E2E prefetch to C3 ({\it i.e.} C4) brings down the prefetch latency to -50ms for PF50 and 110ms for PF90 with a PFR of 1.86. The WER increased to 6.0\% which is still comparable to B1.

\vspace{-0.05in}
\subsection{Two-Pass with Cascaded Encoders}
\vspace{-0.05in}

To further address the quality degradation due to latency related optimization in the 1st-pass, a LAS-based two-pass rescoring model \cite{sainath20streaming, SainathPang19} was firstly applied, which is referred to as T1 in Table~\ref{tbl:latency}. It reduces the WER to 5.3\% with comparable endpointer and partial latency metrics; however, the E2E Prefetch is less effective as it only applies to RNN-T EP model. We then train a Cascaded Encoders model (T2 in Table~\ref{tbl:latency}). In the causal only mode, `T2 Single-pass Casual' performs similar to C4; while with the additional non-causal 2nd-pass at EOQ, `T2 Two-pass' achieves a similar WER as T1 but with much better prefetch latency benefiting from the E2E prefetching. In non-streaming application, we can use `T2 Single-pass Non-causal' to get an even better WER of 4.8\%. The computation cost of T2 is much higher than T1 due to the use of beam-search instead of rescoring. We will further investigate techniques in reducing it in future work. With the two-pass Cascaded Encoders model T2, we can achieve a much better WER and latency (including endpointer, partial and prefetch) comparing to the previous published best E2E system B1.

\vspace{-0.05in}
\subsection{Comparison to Conventional Model}
\vspace{-0.05in}

We also included the performance of a conventional hybrid ASR model (B0 in Table~\ref{tbl:latency}) that consists of a low-frame-rate acoustic model emitting context-dependent phonemes \cite{Golan16} (0.1GB), a 764k-word pronunciation model (2.2GB), a 1st-pass 5-gram language-model (4.9GB). RNN-T models (except for B2) have a better quality and endpointer latency comparing to the conventional system B0, reconfirming the finding in \cite{sainath20streaming}. With FastEmit, we can achieve a better quality with very close partial latency (C3). The -150ms median partial latency (PR50) for B0 mainly comes from the stronger LM trained on large amount of text only data that can potentially hypothesis the full query even before the speech finishes, which E2E models currently lack of. However, even with slightly worse partial latency, E2E systems can still achieve much better prefetch latency with the help of E2E prefetching (C4), which directly affects the total latency of streaming ASRs. With an additional non-casual 2nd-pass (T2 Two-pass), we can further bring down the WER to 5.4\% while still having a faster prefetch than B0.
\vspace{-0.05in}
\section{Conclusions}
\label{sec:concl}
\vspace{-0.05in}

In this work, we further improve the quality and latency of the RNN-T based E2E models for streaming applications. We replaced the encoder LSTM layers with the recently developed \emph{Conformer} layers \cite{gulati2020conformer} for quality and adopted a simple yet effective latency regularization technique, \emph{FastEmit} \cite{yu21fastemit}, which is generic to any transducer based models. Furthermore, we employed a \emph{Cascaded Encoders} \cite{arun21cascade} RNN-T model that shares the same RNN-T decoder for the two passes. It gives us a system that is better and faster than the previous best E2E system \cite{sainath20streaming} and surpassing the conventional model in quality and all latency metrics.
\vspace{-0.1in}

\vfill\clearpage
%\newpage

% References should be produced using the bibtex program from suitable
% BiBTeX files (here: strings, refs, manuals). The IEEEbib.bst bibliography
% style file from IEEE produces unsorted bibliography list.
% -------------------------------------------------------------------------
\bibliographystyle{IEEEbib}
\bibliography{main}

\begin{thebibliography}{10}

\bibitem{sainath20streaming}
T.~N. {Sainath}, Y.~{He}, B.~{Li}, et~al.,
\newblock ``{A Streaming On-Device End-To-End Model Surpassing Server-Side
  Conventional Model Quality and Latency},''
\newblock in {\em Proc. ICASSP}, 2020.

\bibitem{li2020towards}
B.~Li, S.-Y. Chang, T.~N. Sainath, et~al.,
\newblock ``{Towards fast and accurate streaming end-to-end ASR},''
\newblock in {\em Proc. ICASSP}, 2020.

\bibitem{yu21fastemit}
J.~Yu, C.-C. Chiu, B.~Li, et~al.,
\newblock ``{FastEmit: Low-latency Streaming ASR with Sequence-level Emission
  Regularization},''
\newblock in {\em Proc. ICASSP}, 2021.

\bibitem{gulati2020conformer}
A.~Gulati, J.~Qin, C.-C. Chiu, et~al.,
\newblock ``{Conformer: Convolution-augmented Transformer for Speech
  Recognition},''
\newblock in {\em Proc. Interspeech}, 2020.

\bibitem{arun21cascade}
A.~Narayanan, T.~N. Sainath, R.~Pang, et~al.,
\newblock ``{Cascaded encoders for unifying streaming and non-streaming ASR},''
\newblock in {\em Proc. ICASSP}, 2021.

\bibitem{li2020comparison}
J.~Li, Y.~Wu, Y.~Gaur, et~al.,
\newblock ``{On the Comparison of Popular End-to-End Models for Large Scale
  Speech Recognition},''
\newblock {\em arXiv:2005.14327}, 2020.

\bibitem{Ryan19}
Y.~He, T.~N. Sainath, R.~Prabhavalkar, et~al.,
\newblock ``{Streaming End-to-end Speech Recognition For Mobile Devices},''
\newblock in {\em Proc. ICASSP}, 2019.

\bibitem{CC18}
C.-C. Chiu, T.~N. Sainath, Y.~Wu, et~al.,
\newblock ``{State-of-the-art Speech Recognition With Sequence-to-Sequence
  Models},''
\newblock in {\em Proc. ICASSP}, 2018.

\bibitem{Graves12}
A.~Graves,
\newblock ``{Sequence Transduction with Recurrent Neural Networks},''
\newblock {\em CoRR}, vol. abs/1211.3711, 2012.

\bibitem{Chan15}
W.~Chan, N.~Jaitly, Q.~V. Le, and O.~Vinyals,
\newblock ``Listen, attend and spell,''
\newblock {\em CoRR}, vol. abs/1508.01211, 2015.

\bibitem{KimHoriWatanabe17}
S.~Kim, T.~Hori, and S.~Watanabe,
\newblock ``Joint {CTC}-attention based end-to-end speech recognition using
  multi-task learning,''
\newblock in {\em Proc. ICASSP}, 2017.

\bibitem{Jaitly16}
N.~Jaitly, D.~Sussillo, Q.~V. Le, et~al.,
\newblock ``{An Online Sequence-to-sequence Model Using Partial
  Conditioning},''
\newblock in {\em Proc. NIPS}, 2016.

\bibitem{tsunoo2020streaming}
E.~Tsunoo, Y.~Kashiwagi, and S.~Watanabe,
\newblock ``{Streaming Transformer ASR with Blockwise Synchronous Inference},''
\newblock {\em arXiv:2006.14941}, 2020.

\bibitem{colin17}
C.~Raffel, M.~Luong, P.~J. Liu, et~al.,
\newblock ``{Online and Linear-Time Attention by Enforcing Monotonic
  Alignments},''
\newblock in {\em Proc. ICML}, 2017.

\bibitem{fan2018online}
R.~Fan, P.~Zhou, W.~Chen, et~al.,
\newblock ``An online attention-based model for speech recognition,''
\newblock {\em arXiv:1811.05247}, 2018.

\bibitem{moritz2019triggered}
N.~Moritz, T.~Hori, and J.~Le~Roux,
\newblock ``Triggered attention for end-to-end speech recognition,''
\newblock in {\em Proc. ICASSP}, 2019.

\bibitem{wang2020low}
C.~Wang, Y.~Wu, S.~Liu, et~al.,
\newblock ``{Low Latency End-to-End Streaming Speech Recognition with a Scout
  Network},''
\newblock {\em arXiv:2003.10369}, 2020.

\bibitem{SainathPang19}
T.~N. Sainath, R.~Pang, D.~Rybach, et~al.,
\newblock ``{Two-Pass End-to-End Speech Recognition},''
\newblock in {\em Proc. Interspeech}, 2019.

\bibitem{Ding19}
D.~Zhao, T.~N. Sainath, D.~Rybach, et~al.,
\newblock ``{Shallow-Fusion End-to-End Contextual Biasing},''
\newblock in {\em Proc. Interspeech}, 2019.

\bibitem{sainath20emitting}
T.~N. Sainath, R.~Pang, D.~Rybach, et~al.,
\newblock ``Emitting word timings with end-to-end models,''
\newblock in {\em Proc. Interspeech}, 2020.

\bibitem{inaguma2020minimum}
H.~Inaguma, Y.~Gaur, L.~Lu, et~al.,
\newblock ``{Minimum latency training strategies for streaming
  sequence-to-sequence ASR},''
\newblock in {\em Proc. ICASSP}, 2020.

\bibitem{chiu2019comparison}
C.-C. Chiu, W.~Han, Y.~Zhang, et~al.,
\newblock ``A comparison of end-to-end models for long-form speech
  recognition,''
\newblock in {\em Proc. ASRU}, 2019.

\bibitem{wu2018group}
Y.~Wu and K.~He,
\newblock ``Group normalization,''
\newblock in {\em Proc. ECCV}, 2018.

\bibitem{dai2019transformerxl}
Z.~Dai, Z.~Yang, Y.~Yang, et~al.,
\newblock ``{Transformer-XL: Attentive Language Models Beyond a Fixed-Length
  Context},''
\newblock in {\em ACL}, 2019.

\bibitem{wang2019transformer}
Y.~Wang, A.~Mohamed, D.~Le, et~al.,
\newblock ``{Transformer-based acoustic modeling for hybrid speech
  recognition},''
\newblock {\em arXiv:1910.09799}, 2019.

\bibitem{Arun19}
A.~Narayanan, R.~Prabhavalkar, C.-C. Chiu, et~al.,
\newblock ``{Recognizing Long-Form Speech Using Streaming End-to-End Models},''
\newblock in {\em Proc. ASRU}, 2019.

\bibitem{luiz2003web}
L.A. Barroso, J.~Dean, and U.~Hölzle,
\newblock ``{Web Search for a Planet: The Google Cluster Architecture},''
\newblock {\em IEEE Micro}, vol. 23, pp. 22--28, 2003.

\bibitem{jonassen2012prefetching}
S.~Jonassen, B.~B. Cambazoglu, and F.~Silvestri,
\newblock ``Prefetching query results and its impact on search engines,''
\newblock in {\em Proc. SIGIR}, 2012.

\bibitem{chang20low}
S.-Y. Chang, B.~Li, D.~Rybach, et~al.,
\newblock ``Low latency speech recognition using end-to-end prefetching,''
\newblock in {\em Proc. Interspeech}, 2020.

\bibitem{Graves06}
A.~Graves, S.~Fernandez, F.~Gomez, and J.~Schmidhuber,
\newblock ``{Connectionist Temporal Classification: Labeling Unsegmented
  Sequenece Data with Recurrent Neural Networks},''
\newblock in {\em Proc. ICML}, 2006.

\bibitem{Schuster2012}
M.~Schuster and K.~Nakajima,
\newblock ``Japanese and {K}orean voice search,''
\newblock in {\em Proc. ICASSP}, 2012.

\bibitem{Biadsy17}
F.~Biadsy, M.~Ghodsi, and D.~Caseiro,
\newblock ``{Effectively Building Tera Scale MaxEnt Language Models
  Incorporating Non-Linguistic Signals},''
\newblock in {\em Proc. Interspeech}, 2017.

\bibitem{AbadiAgarwalBarhamEtAl15}
M.~{Abadi et al.},
\newblock ``Tensorflow: Large-scale machine learning on heterogeneous
  distributed systems,'' {Available online:
  http://download.tensorflow.org/paper/whitepaper2015.pdf}, 2015.

\bibitem{shen2019lingvo}
J.~Shen, P.~Nguyen, Y.~Wu, et~al.,
\newblock ``Lingvo: a modular and scalable framework for sequence-to-sequence
  modeling,''
\newblock {\em arXiv:2005.08100}, 2019.

\bibitem{KingmaBa15}
D.~P. Kingma and J.~Ba,
\newblock ``Adam: A method for stochastic optimization,''
\newblock in {\em Proc. ICLR}, 2015.

\bibitem{liao2013large}
H.~Liao, E.~McDermott, and A.~Senior,
\newblock ``{Large Scale Deep Neural Network Acoustic Modeling with
  Semi-supervised Training Data for YouTube Video Transcription},''
\newblock in {\em Proc. ASRU}, 2013.

\bibitem{kim2017mtr}
C.~Kim, A.~Misra, K.~Chin, et~al.,
\newblock ``{Generation of Large-Scale Simulated Utterances in Virtual Rooms to
  Train Deep-Neural Networks for Far-Field Speech Recognition in {Google
  Home}},''
\newblock in {\em Proc. Interspeech}, 2017.

\bibitem{Li12}
J.~Li, D.~Yu, J.~Huang, and Y.~Gong,
\newblock ``{Improving Wideband Speech Rcognition using Mixed-bandwidth
  Training Data in CD-DNN-HMM},''
\newblock in {\em Proc. SLT}, 2012.

\bibitem{Golan16}
G.~Pundak and T.~N. Sainath,
\newblock ``Lower frame rate neural network acoustic models,''
\newblock in {\em Proc. Interspeech}, 2016.

\end{thebibliography}

\end{document}